# A Brief Review on Models for Performance Evaluation in DSS Architecture

Ghassem Tofighi, Kaamran Raahemifar, Anastasios N. Venetsanopoulos
Department of Electrical and Computer Engineering
Ryerson University
Toronto, Canada
gtofighi@ryerson.ca, kraahemi@ee.ryerson.ca, tasvenet@ryerson.ca

*Abstract*—Distributed Software Systems (DSS) are used these days by many people in the real time operations and modern enterprise applications. One of the most important and essential attributes of measurements for the quality of service of distributed software is performance. Performance models can be employed at early stages of the software development cycle to characterize the quantitative behavior of software systems. In this research, performance models based on fuzzy logic approach, queuing network approach and Petri net approach have been reviewed briefly. One of the most common ways in performance analysis of distributed software systems is translating the UML diagrams to mathematical modeling languages for the description of distributed systems such as queuing networks or Petri nets. In this paper, some of these approaches are reviewed briefly. Attributes which are used for performance modeling in the literature are mostly machine based. On the other hand, end users and client parameters for performance evaluation are not covered extensively. In this way, future research could be based on developing hybrid models to capture user's decision variables which make system performance evaluation more user driven.

I. INTRODUCTION

Distributed Software Systems (DSS) are used these days by many people in the real time operations and modern enterprise applications. A distributed software system can be considered as a set of sequential processes that compute locally and interact among themselves through communication channels. Examples of distributed software systems are Complex e-commerce sites, data acquisition networks and distributed computing environment are very expensive to develop and maintain. These systems should provide an adequate level of qualities in order to be used. Early quality analysis help to identify and correct issues from the early stages of the software development, in order to compare design alternatives. It is also useful to identify system bottlenecks [1].

One of the most important and essential attributes of measurements for quality of distributed software is performance. Software performance includes many quality factors of the distributed software systems such as software itself, the platform and operating system, middleware, hardware, communication networks, and also client and end users. Commonly, it needs a long and expensive tuning process at the time of product integration and testing. This tuning process occurs too late when software is in its development cycle. In this way, early evaluation of performance is a vital need in software design. Early evaluation applies some kind of model, because the software does not exist in the time of evaluation. It is important because market pressures. There is a huge competition among companies which forces use of performance evaluation models to prove the software quality. Developing and verifying these models is the domain which is called performance analysis.

There are many approaches for modeling performance of a distributed software system. Soft Computing, Queuing Network, Petri net, Pattern-Based, UML Based, Hierarchical Modeling, Component-Based Modeling, Scenario-Based, Software Architecture Analysis Methods (SAAM), Hybrid Approaches such as UML-Petri net, UML-Stochastic Petri net, and Queue Petri Nets are some of well-known approaches in this field.

In this paper, we briefly review some recent researches in the field of performance analysis based on these three approaches:

*A. Fuzzy Logic-based Approach*

These approaches are used when we cannot measure the attributes precisely and we should use some soft computing methods such as fuzzy logic to overcome this problem. These approaches are trying to mimic and emulate the ability of human mind for evaluating performance measures in uncertainty and imprecision environment. It is used in development stage of distributed software system architecture. For example, they are employed to estimate the cost of or effort of software projects when they are described by either numerical data or linguistic values.

*B. Queuing Networks-based Approach*

One of the common modeling paradigms which consists of a set of interconnected queues is this approach. Each queue represents a service station, which serves requests sent by customers. Several models based on queuing networks have been reviewed in this paper. Most of them are based on converting UML diagrams of distributed softwares to queuing networks.

*C. Queuing Networks-based Approach*

A Petri net is a graphical and mathematical modeling language which consists of four basic elements which are places, transitions, tokens and arks. In the similar manner to queuing network approach, in many methods UML diagrams are translated to Petri nets for performance analysis.

In the following sections we introduce some of these methods with more details.

## II. FUZZY LOGIC-BASED APPROACH

In practice, many factors that describe software projects, such as the experience of programmers or the complexity of modules, are measured in terms of an ordinal scale composed of qualifications such as *'low'* and *'high'* [2]. In these situations, measuring the attributes precisely is impossible. Some soft computing methods such as fuzzy logic is employed to overcome this problem. Software performance evaluation models are required which consider imprecision and uncertainty associated with linguistic variables. One of the most important usages of these models is estimation of software development effort by analogy and similarity in early stages of software development. The similarity of two software projects, which are described and characterized by a set of attributes, is often evaluated by measuring the distance between these two projects through their sets of attributes.

In [2] a set of new similarity measures based on fuzzy logic is evaluated. They can be used when the software project attributes are described with linguistic variables such as *'low'* and *'high'*. These measures are also applicable when the variables are numeric while relocating numeric values into a singleton fuzzy set (no uncertainty) or into a fuzzy number (uncertainty). Attributes considered in this model are software size, project mode plus 15 cost drivers.

In [3] the economic evaluation of information system projects is analyzed based on triangular fuzzy numbers. They developed a fuzzy model for evaluating information system projects based on their present value. Three parameters representing three possible values of project costs, benefits, evaluation periods and discount rate are modeled. Figure 1 shows an example of modeling these three attributes using triangular fuzzy numbers.

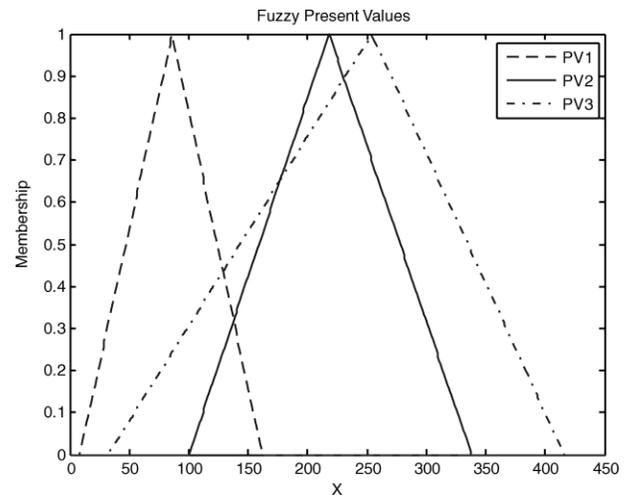

Figure. 1 A plot of the three fuzzy present values [3].

These fuzzy logic-based approaches are commonly used in real-life projects with high degree of uncertainty and risk. However, these models are not comprehensive, because they just consider some performance measures in development phase and they cannot evaluate the performance in production environments.

## III. QUEING NETWORKS-BASED APPROACH

In [1] performance and specification models are integrated to provide a tool for quantitative evaluation of software architecture in the design phase. Attributes considered in this research are number of service centres, service rate of service centre, arrival rate of requests at service centre, umber of servers in service centres, routing procedure of requests, number of request circulating in the system, physical resources available  system workloads, and also network topology.

This approach is for performance modeling of UML software architectures. It considers annotations based on the UML Performance Profile, and derives a performance model based on Queuing Network. Performance evaluation of the QN by efficient algorithms provides a set of steady-state performance indices that characterize the software system behavior. Afterwards, these results are reported back in the UML software specification model. They are reported as tagged values in the diagrams.  Figure 2 represents mapping between UML and performance model elements.

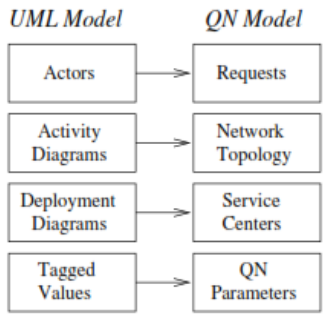

Figure 2: Mapping Between UML and Performance Model Elements [1].

Figure 4 shows an example of annotated Deployment and Activity diagram. Each node represents a processor with the characteristics described with the tagged values.

The QN model of Figure 7 is an example of a network derived from the Deployment and Activity diagrams of Figure 3, assuming that the Activity diagram is associated to an actor stereotyped as <<*OpenUser*>>.

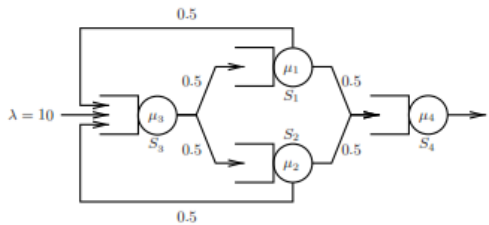

Figure. 3: Corresponding Open QN of annotated Deployment and Activity diagrams [1].

In this example, there is only one job class, which corresponds to the only actor in the UML model. The QN is made of four service centers $S_1,..,S_4$ which correspond to the four resources represented in the Deployment diagram. In this diagram, $S_1$ corresponds to the "Proxy Server" node, $S_2$ corresponds to the "File Server" node, $S_3$ corresponds to the "Workstation" node and $S_4$ corresponds to the "Backup Server" node. The topology of the QN is derived from the Activity diagram. Each UML transition from activity $i$ to activity $j$ is mapped into an edge from server $S_i$ to server $S_j$. The service rates $\mu_1,..,\mu_4$ are derived from the `PArate` tags of the Deployment diagram.

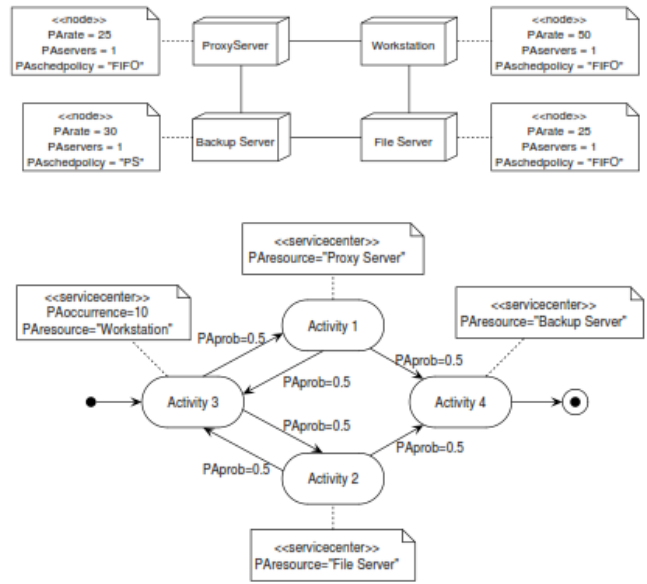

Figure 4: Example of annotated Deployment (above) and Activity (below) diagrams [1].

The analysis of the product-form *QN* of Figure 3 provides a set of average performance indices that include mean number of requests, component utilization and throughput and average response time. Table I shows some numerical results for the specified set of parameters value, for each system component.

TABLE I
Performance Results for the QN of Figure 3: Mean number of Customers ($N_i$), Utilization ($U_i$), Throughput ($X_i$), and and Mean Response Time ($R_i$) [1].

| Resource | $N_i$ | $U_i$ | $X_i$ | $R_i$ |
|---|---|---|---|---|
| Workstation | 0.67 | 0.4 | 20.0 | 0.03 |
| File server | 0.67 | 0.4 | 10.0 | 0.069 |
| Proxy Server | 0.67 | 0.4 | 10.0 | 0.069 |
| Backup Server | 0.6 | 0.33 | 10.0 | 0.05 |

In [4], another approach based on queuing networks models for performance prediction of software systems at the software architecture level, specified by UML. In this approach, attributes considered for performance evaluation are Range of number of clients accessing the system, average think time of each client, number of layers in the software system, relationship between the machines and software components, number of CPUs and disks on each machine, thread limitation, uplink and downlink capacities of the connectors connecting machines running adjacent layers of the system, size of packets of the links, service time required to service one request by a software layer, forward transition probability, rating factors of the CPU and the disks of each machines in the system.

A network-based video server system, where a user requests video frames from a video server through a network connection is presented as a case study for this approach.

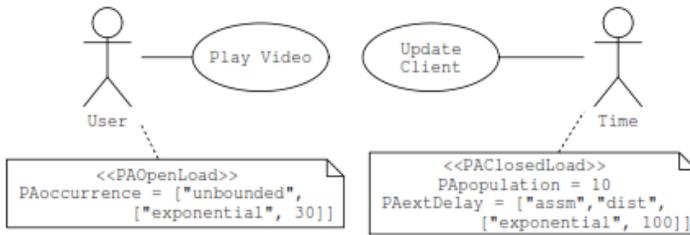

Figure 5: Annotated Use Case diagram [4].

Figure 5 represents an annotated Use Case diagram. There are two actors, User and Time respectively. The first actor (User) represents an open population of requests. It is exponentially distributed interarrival time with mean of 8 time units. The second actor (Time) represents a fixed population of 10 requests which continuously circulate through the system; An exponentially distributed amount of time outside the system is spent for each request. Its mean is 10 time units, before interacting again. Actor User represents the requests which are generated by users requesting videos from the video server. Actor Time represents periodic updates sent by the video server to the player software on the client workstation. Figure 6 shows the corresponding QN for Show Video Activity diagram.

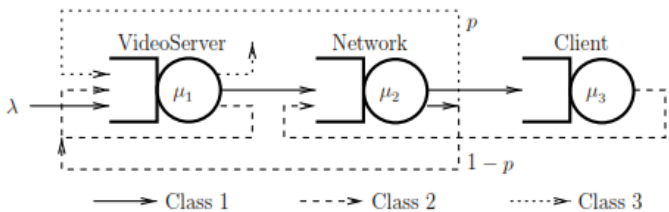

Figure 6: QN for the Show Video Activity diagram [4].

An algorithm for automatic translation of annotated UML specifications into multiclass QN performance models is developed. This approach is one of the most comprehensive approaches which considers both machine-based and also user-based parameters in performance evaluation.

In [5], authors modeled layered software system as a closed Product Form Queuing Network (PFQN) and solve it for finding performance attributes of the system. The attributes considered in this approach is similar to the attributes of [4].

## IV. PETRI NET-BASED APPROACH

Similar to the previous section, many models which are using Petri nets for modeling performance of a distributed software system are developing algorithms for translation from UML activity diagrams to Petri nets. In [10], The UML Profile for Schedulability, Time Specification and Performance allows the specification of quantitative information converts to the UML model directly.

In [7], authors focused on activity diagrams to translate them into GSPN models. They developed stochastic Petri nets model from UML activity diagrams. The attributes they considered for performance evaluation are routing rate, action duration, system response time. In [8], performance evaluation model is developed for Agent-based system using Petri net approach. In this model, system load, system delays, system routing rate, latency of process, and CPU time are considered for performance evaluation. This approach is integrated in the early stages of the software development process. In this way, it is possible to predict the behavior without the need of carrying out the complete implementation phase.

In this method, first system is modeled using with pa-UML diagrams. Afterwards, state transition diagram of the system is created using this model. Figure 7 shows the state transition diagram for the user.

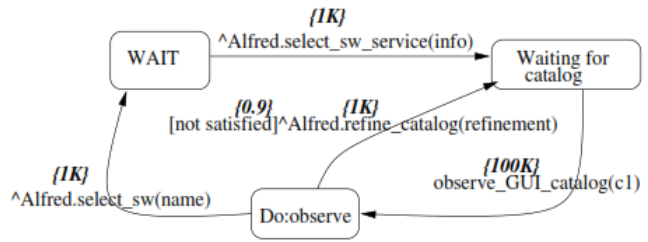

Figure 7: State transition diagram for the user [8].

Petri nets used to model these state transitions because of its significant capabilities. There are also well-known analytic techniques to study system performance in stochastic Petri net models. Figure 8 shows user Petri net component model of the system.

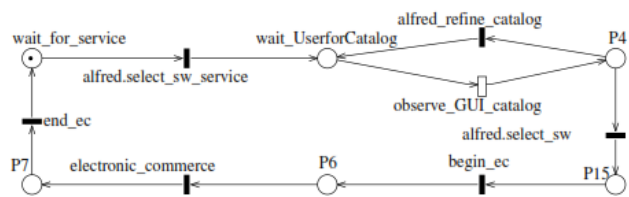

Figure 8: User Petri net component [8].

In [10], network time is the only attribute which is considered for performance evaluation of Internet based software retrieval systems using Petri nets. Finally in [9], Routing rate, action duration, and system response time are considered to translate UML activity diagram into stochastic Petri net model that allows computing performance indices. They have described the kind of annotations suitable to model performance requirements in the context of Activity Diagrams and translation algorithm of them.

## V. Conclusion And Future Works

One of the most important and essential attributes of measurements for the quality of service of distributed software is performance. Performance models can be applied at early phases of the software development cycle. They help to characterize the quantitative behavior of software systems.

Parameters models for performance evaluation are mostly machine based such as CPU usage, Network time, RAM size, message size. These models are implemented at the early stage of the software life cycle. Among the papers reviews in this research, [4] is one of the most comprehensive approaches which considers both machine-based and also user-based parameters in performance evaluation.

End users and client parameter for performance evaluation are not covered in most of other modeling approaches. In terms of end users, lots of uncertain and imprecision parameters exist. Hence, uncertainties and imprecision of parameters should also be considered.

The future research in the field of performance analysis for distributed software systems could be based on developing hybrid models which capture user's decision variables that make system performance evaluation to be more users driven. In this way, it will be a complement to the existing models which are mostly are based on machine parameters.